\documentclass[nonacm,11pt]{acmart}
\usepackage[final]{changes}
\usepackage{hyperref}

\usepackage{graphicx}
\usepackage{url}

\usepackage[title]{appendix}
\usepackage{bytefield}
\usepackage{xspace}
\usepackage{enumitem}
\usepackage{multirow}
\usepackage{color}
\usepackage[noend]{algpseudocode}
\usepackage{algorithm}
\usepackage{algorithmicx}
\algdef{SE}[DOWHILE]{Do}{doWhile}{\algorithmicdo}[1]{\algorithmicwhile\ #1}%
\usepackage[nolist,nohyperlinks]{acronym}
\graphicspath{{figures/}}

\newcommand{\vfour}{IPv4\xspace}
\newcommand{\vsix}{IPv6\xspace}
\newcommand{\icmpsix}{ICMPv6\xspace}
\newcommand{\eui}{EUI-64\xspace}
\newcommand{\edgy}{edgy\xspace}
\newcommand{\Edgy}{Edgy\xspace}
\newcommand{\edge}{\replaced{periphery}{edge}\xspace}
\newcommand{\ehost}{endhost\xspace}
\newcommand{\ehosts}{endhosts\xspace}
\newcommand{\lhop}{last hop\xspace}
\newcommand{\lhops}{last hops\xspace}
\newcommand{\yarrp}{yarrp\xspace}
\newcommand{\eg}{e.g.,\xspace}
\newcommand{\ie}{i.e.,\xspace}
\newcommand{\etal}{et al.\xspace}

\begin{document}

\title{Discovering the IPv6 Network \replaced{Periphery}{Edge}}

\author{Erik C. Rye}
\orcid{0000-0002-8151-8252}
\affiliation{%
  \institution{CMAND}
}
\email{rye@cmand.org}
\author{Robert Beverly}
\orcid{0000-0002-5005-7350}
\affiliation{%
  \institution{Naval Postgraduate School}
}
\email{rbeverly@nps.edu}

\begin{abstract}
We consider the problem of discovering the \vsix
network \edge, \ie the \lhop router connecting \ehosts in the
\vsix Internet.  Finding the \vsix \edge using active probing is
challenging due to the \vsix address space size, wide variety of
provider addressing and subnetting schemes, and incomplete topology
traces.
As such, existing topology mapping systems
can miss the large footprint of the \vsix \edge, disadvantaging
applications ranging from \vsix census studies to geolocation and
network resilience.  We introduce ``edgy,''
an approach to explicitly discover the \vsix network \edge, and use it to
find $>$~64M \vsix \edge router addresses and $>$~87M links to
these last hops -- several
orders of magnitude more than in currently available \vsix topologies.
Further, 
 only 0.2\% of edgy's discovered addresses are known to existing
 \vsix hitlists.

\end{abstract}

\maketitle

\section{Introduction}
\label{sec:intro}

Among the unique properties inherent to \vsix's large address space
size are ephemeral and dynamic addressing, allocation
sparsity and diversity, and a lack of address translation.  These well-known
properties complicate efforts to map the \added{infrastructure} topology of the \vsix
Internet.  
While previous research has tackled problems of target
selection, speed, and response rate-limiting in active \vsix topology
probing~\cite{imc18beholder}, the \vsix \emph{\edge} \added{-- \lhop 
routed infrastructure connecting end hosts -- } is challenging to
discover, and difficult to discern.

Discovery of the \vsix \edge is important not only to the
completeness of network topology mapping, but provides a crucial
supporting basis for many applications.  For instance, \vsix
adoption~\cite{Czyz:2014:MIA:2740070.2626295,Zander:2018:WYI:3185332.3158374,pujol2017understanding}, 
census~\cite{Plonka:2015:TSC:2815675.2815678}, and reliability and outage
studies~\cite{Luckie:2017:IRO:3098822.3098858} all depend in
part on a complete and accurate map of the \vsix topology inclusive
of the \edge, while 
understanding provider address allocation policies and utilization
also requires completeness~\cite{Foremski:2016:EUS:2987443.2987445,v6exhaust-ic16}.  
Similarly, work on \vfour to \vsix network
congruence~\cite{Dhamdhere:2012:MDI:2398776.2398832,livadariu2015leveraging} and
\vsix geolocation~\cite{imc13dns} can utilize 
\vsix topologies.
Further, our work illuminates 
potential security and privacy vulnerabilities inherent in 
the way today's \vsix \edge is deployed~\cite{czyz2016,rye2019eui64}. 

We present ``\edgy,'' a new technique to explicitly 
discover the \vsix \edge.  In contrast to \vsix
scanning~\cite{Murdock:2017:TGI:3131365.3131405,rfc7707}, passive
collection~\cite{Plonka:2015:TSC:2815675.2815678}, or 
hitlists~\cite{Gasser:2018:CEU:3278532.3278564,fan2010selecting}, which, by
construction, target \ehosts, 
\edgy is specifically designed to find \lhop routers
and subnetworks in the \vsix Internet.
Our contributions include:
\begin{enumerate}
  \item Edgy, an algorithm to discover, identify, and enumerate the
   \vsix \edge.
  \item Active measurement using \edgy to find
        64.8M \lhop router addresses
        and 87.1M edges to these \lhops from a single vantage.
  \item Discovery of \edge addresses that are 99.8\% disjoint from 
         current \vsix hitlists~\cite{Gasser:2018:CEU:3278532.3278564}
         and orders of magnitude larger than existing \vsix
         topology snapshots~\cite{caida-topov6}, suggesting that
         edgy is complementary to these prior approaches.
  \item Discovery of 16M EUI-64 \lhop addresses, suggesting a potential
        vulnerability to security and privacy. 
\end{enumerate}


\section{Background and Related Work}
In this work, we define the ``\edge'' not to be servers or clients,
but rather the \lhop router connecting network \ehosts.
Whereas significant prior work has developed techniques for
\vsix \ehost
discovery~\cite{Murdock:2017:TGI:3131365.3131405,rfc7707,Gasser:2018:CEU:3278532.3278564}, 
comparatively little work has
explored the \vsix \edge.  

The large address space in \vsix removes the
need for address translation; thus, while many \vfour hosts are
connected via NATs~\cite{rfc2663}, the \vsix \edge typically extends into customer
premises.  Indeed, in \vsix, the \ac{CPE} is a
router, implying that in conjunction with the rapid increase in \vsix
adoption~\cite{Czyz:2014:MIA:2740070.2626295,Zander:2018:WYI:3185332.3158374}, 
the \vsix \edge is considerably larger than in \vfour, especially
for residential networks. 

\begin{figure}[t]
 \centering
 \resizebox{0.6\columnwidth}{!}{\includegraphics{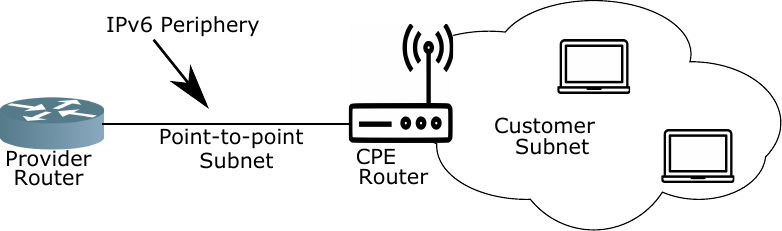}}
 \vspace{-3mm}
 \caption{Common IPv6 architecture: 
an \vsix subnet is assigned to the link 
between the provider and \lhop CPE routers. There is
no NAT or private addressing; a separate
distinct routed IPv6 subnet is assigned to devices attached to the 
\lhop CPE.}
 \vspace{-3mm}
 \label{fig:cpe}
\end{figure}

Figure~\ref{fig:cpe} shows an example of the \vsix \edge we attempt to
discover.  Here,
the point-to-point subnet between the provider and the CPE is assigned
a public \vsix prefix; the subnet on the other side of the CPE
(\eg in the customer's home) 
is also a publicly-routed prefix.  While this example shows a common
residential \vsix architecture, similar designs exist in the
enterprise.

Consider an \vsix traceroute to a random address within a provider's
globally advertised BGP prefix, such as is routinely performed by
existing production topology mapping systems~\cite{caida-ark}.  The
traceroute (Figure~\ref{fig:trace}): i) is unlikely to hit the prefix allocated to a customer
CPE or her network; ii) is even less likely to reach a host within the
customer's network; and iii) does not illuminate the scope,
characteristics, or breadth of subnets within the prefix.  When a
traceroute does not reach its target destination it
is ambiguous: does the last responsive hop
belong to the core of the network, or the \edge?

Passive techniques suffer similar problems in revealing the network
\edge.  For instance, BGP, by design aggregates routes such that the
aggregate visible in a looking glass does not reveal the subnets
within.  And, while there has been significant prior work in
characterizing the \vsix address space, these primarily focus on 
\ehosts.  For example, Plonka and Berger examine and analyze the
addresses and behaviors of \vsix clients connecting to a large
CDN~\cite{Plonka:2015:TSC:2815675.2815678}.  However, this 
passive collection of client requests alone does not reveal the
network \edge on the path to those clients.  

\begin{figure}[t]
 \begin{verbatim}
traceroute to 2a03:4980:2b6:9624:8643:b70f:adae:4f40
    . . . 
 5  2001:7f8:1::a502:4904:1  16.862 ms
 6  2a03:4980::6:0:2  25.948 ms
 7  2a03:4980::b:0:5  39.560 ms
 8  *
 9  *
 \end{verbatim}
 \vspace{-8mm}
 \caption{Randomly
          chosen trace targets are unlikely to discover subnets
          within a prefix, or to elicit
          a response.  It is thus
          ambiguous whether hop 7 is \replaced{a periphery address}{an edge} in this example,
          even though the trace reaches into the destination's /32.}
 \label{fig:trace}
 \vspace{-3mm}
\end{figure}

\section{Methodology}

Our work seeks to perform active probing in a way 
that elicits responses from the \lhop \vsix \edge, rather than network
core infrastructure, servers or other \ehosts.  
Enumerating \lhop router 
addresses, \eg CPE, and inferring networks beyond the \lhops are the
principal goals of edgy. 

Edgy is divided into an initialization stage, followed by active probing
that proceeds in rounds.  
Results from one round of probing are used to
guide probing in subsequent rounds.  
This
section describes edgy;
the complete algorithm is given in
\ref{app:alg}.

\subsection{Edgy}

Because of the massive size of the \vsix address space, edgy relies on
an input set of ``seed traces'' to focus and guide its discovery.  Thus, the
ability of edgy to discover the network \edge depends strongly on the
input seed traces it uses.  
In
\S\ref{sec:method:input} we describe two specific realistic seed
traces
we utilize: i) BGP-informed; and ii) hitlist-informed.

Algorithm~\ref{alg:discoverinit} describes edgy's initialization stage.
Edgy iterates through the input seed and examines the last
responsive hop in each trace, regardless of whether a sequence of same last IP
responses or loops occur.
It maintains the set of targets
that, when used as the traceroute destination, had a given \lhop.
Edgy then finds \emph{unique} last hops -- those that were only
discovered by probing destinations that reside within a single /48 prefix.  The intuition
is to find candidate /48 prefixes that
are likely to be subnetted, and hence contain \edge routers.  By
contrast, if there are two or more probes to targets in different /48s that
elicit the same \lhop, those /48s are less likely to be subnetted, or traces to
targets in these /48s are unresponsive beyond the middle of the network. In
either case, \edgy terminates exploration of these target /48s rather than
continuing to probe them. 

\begin{figure}[t]
 \centering
  \begin{minipage}{.45\textwidth}
    \centering
 \resizebox{1\columnwidth}{!}{\includegraphics{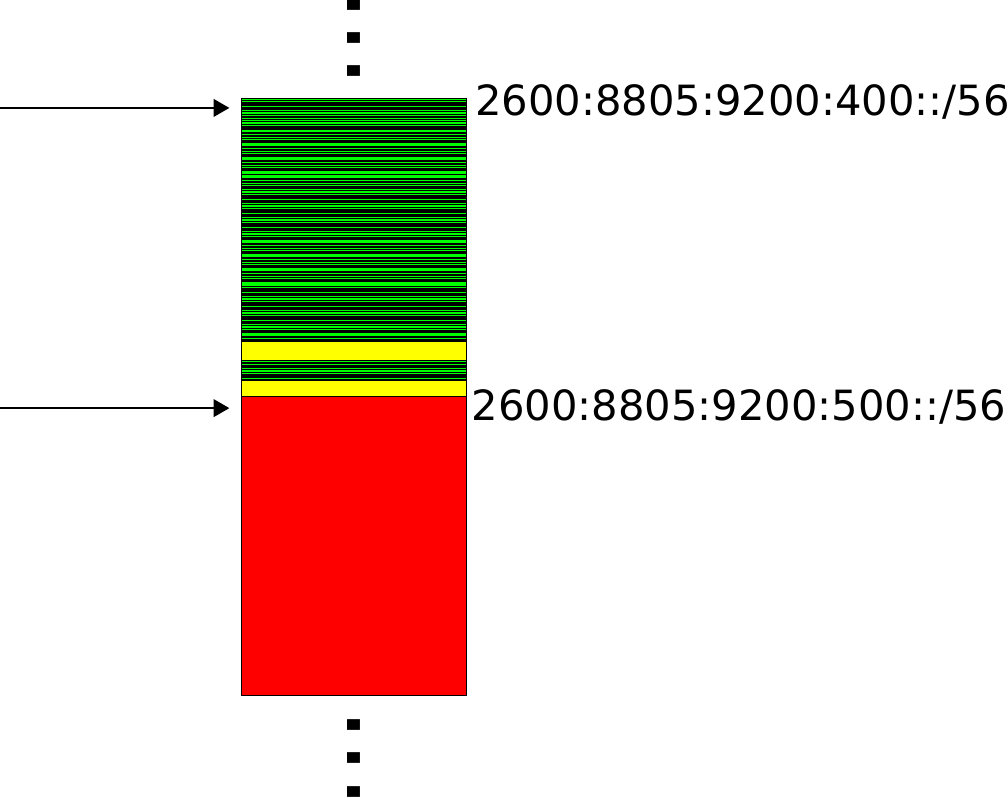}} 
\vspace{-3mm}
    \caption{\replaced{A portion of a target /48 (2600:8805:9200::/48) is shown;
    colors correspond to the true delegated customer subnet sizes that \edgy discovers. Green represents /64, yellow
    /60, and red /56. In the first probing round, \edgy sends probes to each /56
    in a target /48 (represented by arrows).}{\Edgy sends probes 
    to each /56 in a
    target /48 in the first round. Green represents /64s, yellow /60s, and red /56s allocated in a
    Cox /48 prefix.}}
 \label{fig:cox1}
\vspace{-2mm}
  \end{minipage}%
  \hspace{.03\textwidth}
  \begin{minipage}{.45\textwidth}
    \centering
 \resizebox{1\columnwidth}{!}{\includegraphics{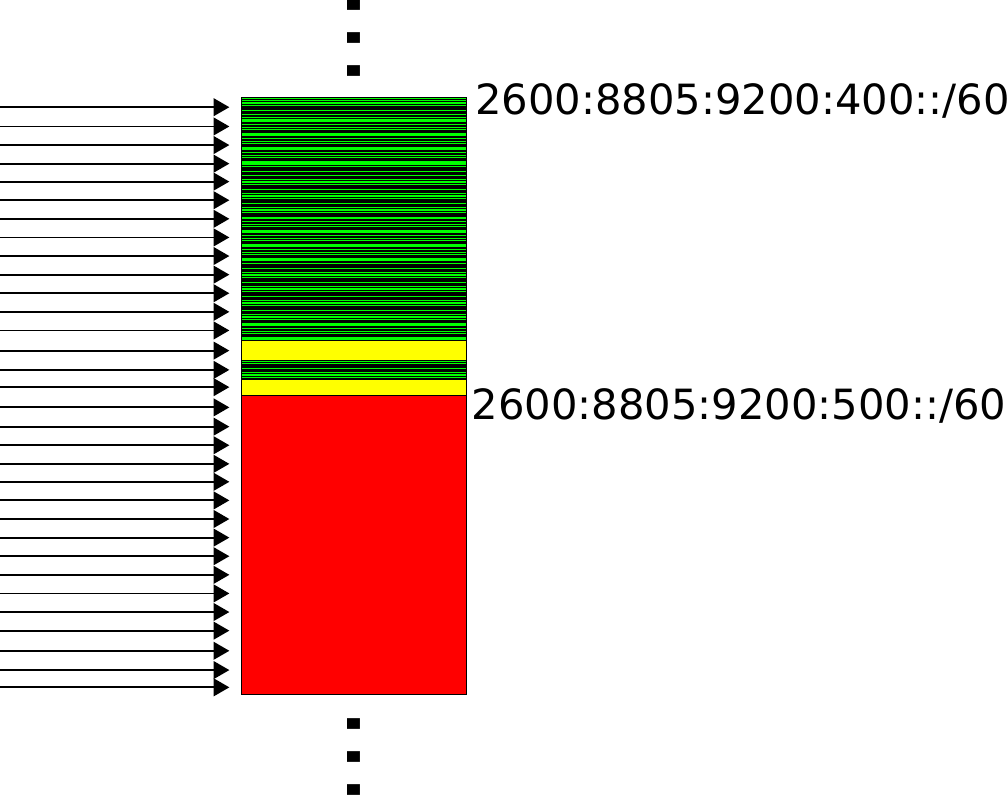}}
\vspace{-3mm}
    \caption{In the second round, probes are sent to each /60 \added{in the
    target /48}. New addresses are discovered in the upper half of this
    \replaced{portion of the target address space where subnet allocation is
    finer- grained}{address space}, but not in the lower
    half. Many operators mix allocation sizes within the same /48.} 
 \label{fig:cox2}
\vspace{-2mm}
  \end{minipage}
\end{figure}

These candidate target /48
prefixes are fed to Algorithm~\ref{alg:discover} 
which 
probes targets within the input prefixes 
at progressively finer
granularities until a stopping condition (a discovery threshold
$\eta$) is reached.
A random \ac{IID} \added{(the 64 least significant bits in an IPv6 address)} for each target subnet is used as the trace
destination.  
Figure~\ref{fig:cox1} depicts an illustration of
edgy's first round behavior targeting an example /48 belonging to Cox
Communications.
%

The first subnet discovery round probes different /56 prefixes 
and serves as a coarse filter to
determine which candidate /48s exhibit an appreciable amount of subnetting and
merit further probing. /56s are used initially 
as~\cite{bcop-prefix} recommends this as a potential subnet size 
for residential customers; therefore, if a /48 is allocated entirely to
residential customers with /56s, the initial probing round should discover 
all of the /56 allocations.
\added{We note, however, that these prefix delegation boundaries are not
mandatory, that it is impossible to know \emph{a priori} what prefix delegation
strategy a provider has chosen, and that networks can be subdivided in a
non-uniform manner for allocations to customers.}
If the number of distinct \lhops found during a probing round
exceeds the threshold $\eta$, 
we further subdivide responsive prefixes for additional 
probing in the next round.
\ifdefined\dox
\ref{app:eta} discusses sensitivity
and choice of $\eta$.
\else
The choice and sensitivity of $\eta$ are discussed in~\cite{rye2020discovering}.
\fi


It has been shown that aliased networks are common in the \vsix Internet,
where every address within a prefix is responsive
despite no actual host being present.
We remove \lhops equal to the probe target,
as well as networks and addresses 
present in the 
publicly curated list of aliases
from Gasser \etal~\cite{Gasser:2018:CEU:3278532.3278564}.  
In addition, we remove replies from
non-routable prefixes -- we observe site- and link-local addresses that fall into
this category -- as well as \vfour-in-\vsix addresses and replies that appear to
be spoofed. 

\added{After removing aliases and bogus replies, target /48s that
generate $>\eta$ unique \lhop addresses proceed to the second round of probing.
In the second round, \edgy sends probes to addresses within each /60 of the
target /48.  Figure~\ref{fig:cox2} depicts an illustration of edgy's second
round behavior, again for the same Cox Communications /48. 
Target /48 networks that generate $>\eta$ unique \lhop addresses
(exclusive of aliases)
move to the next round. The third probing round
sends probes to a random \vsix address in each /62 of the target networks.
Finally, target /48s that exhibit subnetting beyond the /60 level (as evidenced
by four unique \lhops for each /62 within any /60), are probed
at /64 granularity.}

Note that, during testing, we initially explored other \edge discovery
mechanisms. For instance, intuitively, a binary-tree discovery
process that bisects prefixes and probes each half would
programmatically explore subnets.  Unfortunately, such an
efficient
approach performs poorly as providers do not allocate subnets 
uniformly.  In this case, a core router can falsely appear as the common
\lhop for destinations in a common prefix, even when
significant subnetting is present. \added{Additionally, the third round of
probing was added to limit time spent probing target networks at the /64
granularity to those proven to subnet within the final nybble of the network
prefix.}

\subsection{Edgy Input}
\label{sec:method:input}

Edgy takes as input a seed set of traces.  These seed traces are
created from running traceroutes to corresponding seed targets.  We
consider two realistic potential seed target lists: BGP-informed and
hitlist-informed.  The BGP-informed targets assume no prior knowledge
other than global BGP advertisements.  Since BGP routes are readily available
from looking glasses, this scenario is easily replicated by anyone and
models what CAIDA uses to inform their probing.  In our experiments,
we utilize publicly available BGP-informed seed traces collected as
part of an August 2018 effort to uniformly probe every /48 in the \vsix
Internet~\cite{v6exhaust-ic16,caida-routed48}.  
Herein, we term this trace seed
data as the \emph{BGP-informed} seed.

Second, we consider a target set informed by prior knowledge in
the form of passive traces, server logs, or hitlists.  In our
experiments, we utilize a publicly available \vsix hitlist~\cite{Gasser:2018:CEU:3278532.3278564}
that was used to generate a seed set of hitlist-informed
traces~\cite{imc18beholder}.  Herein, we term this trace seed
the \emph{hitlist-informed seed}.

\subsection{Limitations}

There are several potential complications that edgy may 
encounter, and corresponding limitations of our approach and
evaluation.  First, during probing, we depend on receiving a response
from the penultimate traceroute hop along the data path to a destination.
However, the last responsive hop may instead be a different router due
to filtering, loss, or rate-limiting, \ie if the \lhop remains
anonymous.  
This case does not cause false inferences of \edge addresses, but
instead causes edgy to terminate probing of a prefix prematurely. 

Second, we do not have ground-truth in order to determine whether the
\edge we discover \replaced{is}{are} indeed the \lhop before a destination \ehost.
\replaced{While various, and at times conflicting, guidance exists regarding
the size of delegated prefixes~\cite{rfc3177,rfc6177,rfc7381}}{However, as /64 prefixes are (historically) recommended and used
as individual
customer \vsix prefixes~\cite{rfc3177,rfc6177},} discovery of unique /64s is strongly
indicative of discovering \replaced{the periphery}{an edge}.  Additionally, the \edge addresses
we find are frequently formed using EUI-64 addresses where we can 
infer the device type based on the encoded MAC address (see
\S\ref{sec:results:eui64}).
These MAC addresses specifically point to CPE. 
Further, we examine several
metrics of ``edginess'' to better understand the results in
\S\ref{sec:results:edginess}. In particular, we determine whether traces enter
their target network and, if so, quantify how far inside the target network they
reach. We also analyze the \lhop addresses \edgy discovers in order to
understand how many also appear as intermediate hops to different targets. As
intermediate hops, such addresses are unlikely to exist in the \vsix periphery. 

\subsection{Probing}

Probing consists of sending hop-limited \icmpsix
packets; we used the high-speed
randomized \yarrp topology prober~\cite{imc16yarrp} due to the large
number of traces required during edgy's exploration, as well as to
minimize the potential for \icmpsix rate limiting (which is mandated
and common in \vsix~\cite{imc18beholder}).

We use \icmpsix probes as these packets are designed for diagnostics and
therefore are less intrusive than UDP probes.  Further, we send at a
conservative rate while \yarrp, by design, randomizes its probing in
order to minimize network impact.  Last, we follow best established
practices for performing active topology probing: we coordinated with
the network administrators of the vantage point prior to our
experiments and hosted an informative web page on the vantage point
itself describing the experiment and providing opt-out instructions. \added{We
received no opt-out requests during this work.}

\section{Results}
\label{sec:results}


From Sept. to Oct.\  2019 we ran edgy from a
well-connected server in
Lausanne, Switzerland. 
Edgy used \yarrp 
at 
less than
10kpps with the neighborhood
TTL setting to reduce load on routers within five hops of the
vantage point.


%

\subsection{BGP-Informed Seed Results}
\label{sec:gvaresults}

Initializing edgy with the BGP-informed seed data yielded 130,447
candidate
/48 prefixes.
Following Algorithm~\ref{alg:discover},
edgy traced to a random \ac{IID} in each of the 256 constituent /56 subnets in each
/48s (a total of 33,394,432 distinct
traces). 

This first round of probing 33.4M targets discovered 4.6M unique, non-aliased \lhop \vsix
addresses residing in 33,831 distinct /48 prefixes (Table~\ref{table:passive_results}).
Often, the \lhop address is not
contained within the target /48 prefix but in a different /48 prefix belonging
to the same \ac{AS}.
Further, probing different target /48 prefixes in round one resulted
in \lhops within the same /48 (but different than the target /48).
This phenomenon of a many-to-one relationship between the target
prefix and the \lhop prefix persists across rounds as the probing
granularity increases.

The density of discovered \lhop addresses across target
prefixes is non-uniform: 
nearly 75\% 
of the targeted /48 prefixes produce
16 or fewer distinct \lhops. The prefixes in which the \lhops
reside is also highly non-uniform. Of the 33,831 /48s in which \lhop
addresses reside, 11,064 were responsible for only a single
\lhop
address. 
This is likely indicative of a /48 allocation to an end site. On the other
end of the spectrum, a single /48 (\texttt{2001:1970:4000\added{::/48}}) contained over
200,000 unique \lhop addresses.
\texttt{2001:1970:4000::/48} was the \lhop prefix in traces to 1,008 distinct
/48 target prefixes, the most extreme example of many target /48s mapping to a
single \lhop prefix. 

Because a /48 prefix entirely subnetted into /52s should exhibit 16 distinct
\lhops,
we choose $\eta=16$ empirically
as a baseline indication of more granular subnetting.
\ifdefined\dox
We additionally conduct sensitivity
testing with respect to $\eta$, and discuss the choice of $\eta=16$ in
\ref{app:eta}.
\else
The choice and sensitivity of $\eta$ are discussed in detail in
\cite{rye2020discovering}.
\fi

\begin{table}[t]
\centering
  \caption{BGP and Hitlist-Informed Routable Address Discovery by Round}
  \label{table:passive_results}
\vspace{-2mm}
\scriptsize{
\begin{tabular}{|p{10mm}|p{12mm}|p{14mm}|p{13mm}|p{15mm}|
                         p{12mm}|p{14mm}|p{13mm}|p{15mm}|
}
\hline
  & \multicolumn{4}{|c|}{\textbf{BGP-Informed}} & \multicolumn{4}{|c|}{\textbf{Hitlist-Informed}} \\\hline
  \textbf{Round}& \textbf{Prefixes Probed} & \textbf{Unique Last Hops} &
  \textbf{Unique Last Hop /48s} & \textbf{Cum. Unique Last Hops}
  & \textbf{Prefixes Probed} & \textbf{Unique Last Hops} &
  \textbf{Unique Last Hop /48s} & \textbf{Cum. Unique Last Hops}
  \\ \hline
  1 (/56) & \multicolumn{1}{|r|}{130,447} & \multicolumn{1}{|r|}{4,619,692} & \multicolumn{1}{|r|}{33,831} & \multicolumn{1}{|r|}{4,619,692} 
      & \multicolumn{1}{|r|}{111,670} & \multicolumn{1}{|r|}{9,217,137} & \multicolumn{1}{|r|}{89,268} & \multicolumn{1}{|r|}{9,217,137} \\ \hline
  2 (/60) & \multicolumn{1}{|r|}{34,520} & \multicolumn{1}{|r|}{12,228,916} & \multicolumn{1}{|r|}{26,082} & \multicolumn{1}{|r|}{13,410,601}  
      & \multicolumn{1}{|r|}{67,107} & \multicolumn{1}{|r|}{11,021,329} & \multicolumn{1}{|r|}{74,302}  & \multicolumn{1}{|r|}{11,365,910} \\ \hline
  3 (/62) & \multicolumn{1}{|r|}{12,014} & \multicolumn{1}{|r|}{14,770,061} & \multicolumn{1}{|r|}{11,675} & \multicolumn{1}{|r|}{24,832,391}  
      & \multicolumn{1}{|r|}{4,462} & \multicolumn{1}{|r|}{5,428,992} & \multicolumn{1}{|r|}{19,942} & \multicolumn{1}{|r|}{15,569,221} \\ \hline
  4 (/64) & \multicolumn{1}{|r|}{2,641} & \multicolumn{1}{|r|}{15,326,298}  & \multicolumn{1}{|r|}{7,833}  & \multicolumn{1}{|r|}{37,169,357}  
      & \multicolumn{1}{|r|}{1,531} & \multicolumn{1}{|r|}{15,340,591} & \multicolumn{1}{|r|}{32,718} & \multicolumn{1}{|r|}{29,248,703} \\ \hline
\end{tabular}
 }
\vspace{-4mm}
\end{table}

34,520 of the input 130,447 /48 target prefixes passed the $\eta$
threshold in round one.  Each of these /48 prefixes were then 
probed at a /60 granularity  
(4,096 probes to each /48).
Edgy discovers significantly more unique non-aliased \lhop addresses in this
round, $\sim$12.2M,
as the probing is focused on known address-producing target
subnetworks identified in the first round. 

To select target /48s for round three, we use
$\eta=256$ as an indicator of subnetting at a 
granularity finer than /56.
12,014 /48s meet this criteria, and were used as targets
for probing at the /62 granularity 
($\sim$196.8M traces).

Round three, while probing $<10$\% of the input
target seed prefixes, is focused on those with fine-grained
subnetting and helps to
expose subnetting strategies. As the IETF now discourages\replaced{, but
does not forbid, /64 or more-specific subnetting}{subnets more specific than
/64s}~\cite{rfc6177}, 
we are interested in the prevalence of fine-grained subnetting, but
must balance inferring this delegation behavior with probing load.
Because subnetting generally occurs on nybble
boundaries~\cite{rfc6177}, 
by probing /62s, we
are able to detect when target prefixes are subnetted beyond /60s, which is an
indication that perhaps the operator is allocating /64 subnets.  The /62
probing round produced $\sim$14.7M unique \lhop addresses.

The final round is designed to \replaced{enumerate last hop addresses for /64
subnets.}{discover target /48s that contain  
/64 subnets.} 
Edgy selects any prefix with $\eta=4$ prefix-unique 
\lhops within a /60 (because we probe each /62, each /60 contains four targets).
We surmise that four prefix-unique \lhops is an indication that either the
operator subnets at the /62 level, or is assigning /64 networks \deleted{as the interior
network} to their customers.\deleted{in violation of best current practice} The final /64
probing round discovered 15.3M distinct \vsix addresses through exhaustive
probing of 2,641 /48 target prefixes that met the $\eta$ threshold to
be in round four. 

\begin{figure}[t]
 \centering
  \begin{minipage}{.45\textwidth}
    \centering
 \resizebox{1.\columnwidth}{!}{\includegraphics{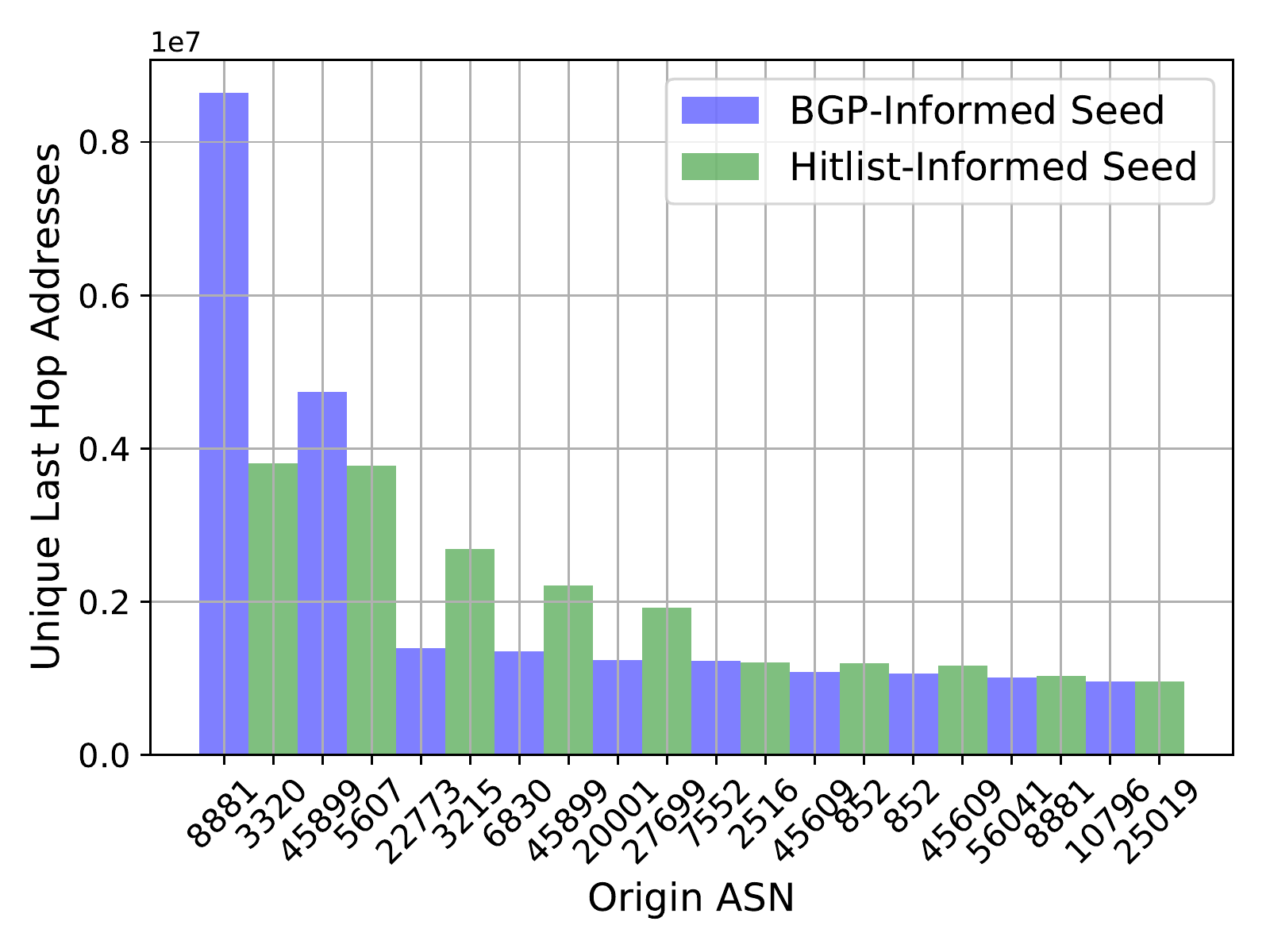}} 
\vspace{-6mm}
    \caption{Top 10 Last Hop ASN}
 \label{fig:top10asn}
\vspace{-4mm}
  \end{minipage}%
  \hspace{.03\textwidth}
  \begin{minipage}{.45\textwidth}
    \centering
 \resizebox{1\columnwidth}{!}{\includegraphics{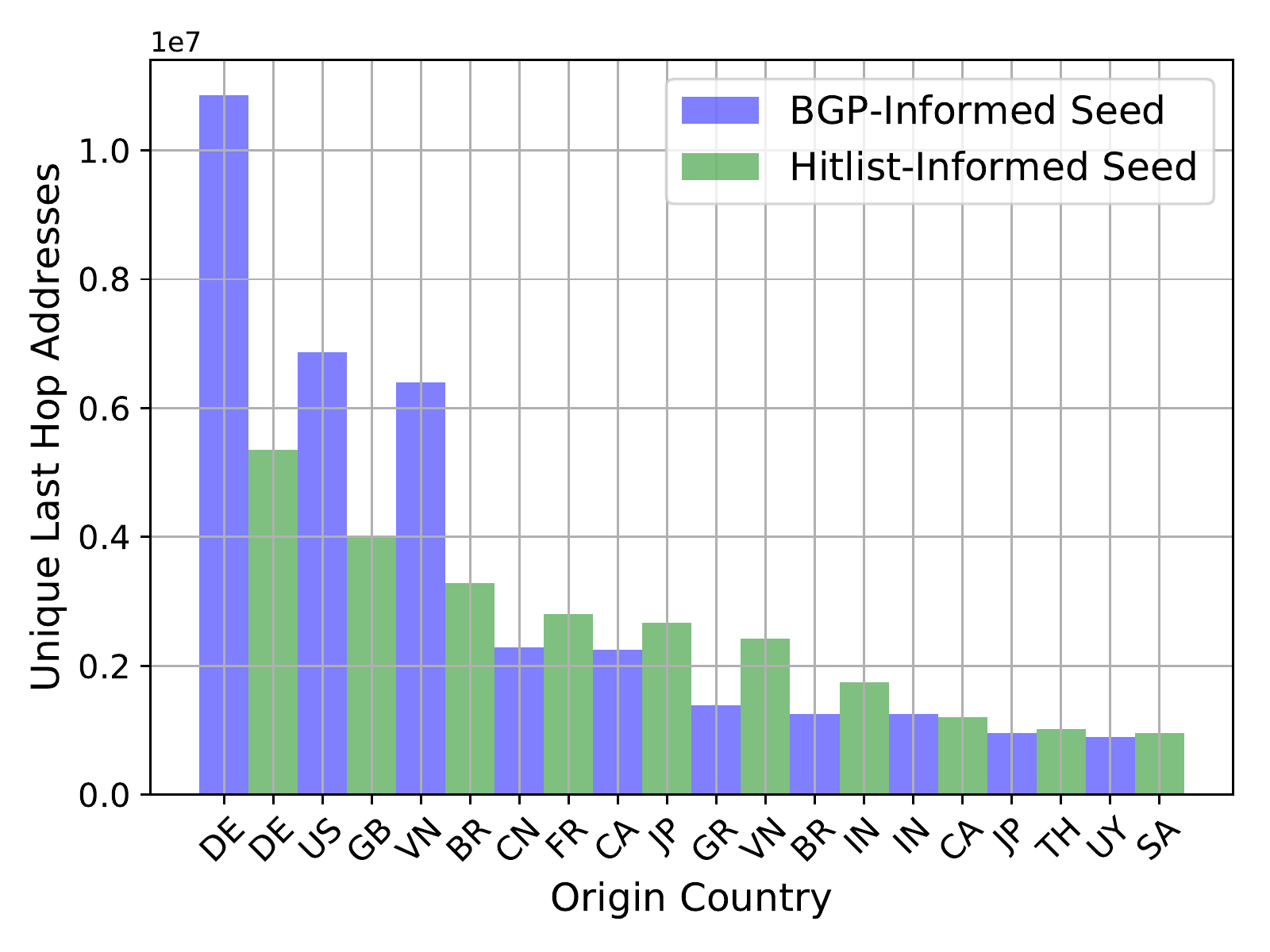}}
\vspace{-6mm}
  \caption{Top 10 Last Hop Country}
 \label{fig:top10country}
\vspace{-4mm}
  \end{minipage}
\end{figure}

Cumulatively, \edgy discovers 
more than 37M
distinct \vsix \lhop addresses 
using the BGP-informed seed.
Table~\ref{table:passive_results} quantifies discovery across
probing rounds. 3,989 \acp{AS}
are represented in the \lhop addresses, corresponding to 143
countries, as reported by Team Cymru's IP to ASN service~\cite{cymru2008ip}.
Figures~\ref{fig:top10asn} and \ref{fig:top10country} summarize the ASes and countries that produced
the largest number of \edge \lhop addresses.

\subsection{Hitlist-Informed Seed Results}
\label{sec:beholderresults}

We replicate the experiment described in \S\ref{sec:gvaresults} seeded with the
hitlist-informed seed traces (from~\cite{imc18beholder}).
Table~\ref{table:passive_results}
shows the per-round results for both the BGP-informed and hitlist-informed seeds.
Algorithm~\ref{alg:discoverinit} on this input seed 
yielded
111,670 target /48 prefixes,
about 20k fewer than the BGP-informed seed. However, the initial /56
probing round discovered nearly twice as many unique \lhop addresses.
The hitlist-informed seed led to almost
double the number of target prefixes in the /60 round as compared 
to the BGP-informed seed,
but discovered nearly 1M fewer \lhops.  
As a result, only 4,462 /48 target prefixes were
probed in the /62 probing round, discovering 5.4M \lhops from
19,942 /48 prefixes. 1,531 target /48s were exhaustively probed 
at the /64 granularity in the fourth round, about 1\% of the input
hitlist seed prefixes. The /64 probing round discovered
over 15M
unique \lhops, indicating that the 1,500 target /48s
each contributed about 10,000 unique addresses on average. We attribute the differences between the
BGP-informed and hitlist-informed seed data results to differences in how the original
source data was collected. For example, the BGP-informed seed data was derived from a
uniform sweep of the advertised \vsix space, while the hitlist-informed seed data
derived from a measurement campaign aimed at networks known to be dense in
customers.

In total, \edge on the hitlist-informed seed discovers over 29M unique
\lhop router
addresses. Nearly half of those addresses
are found in the /64 probing round, during which \edgy exhaustively 
probes all of the /64s in 1,531 /48 target
prefixes.
This suggests that a small number of prefixes have fine-grained
subnetting, and that
substantial \edge topology can be gained by probing a carefully selected set of
target prefixes. Figures~\ref{fig:top10asn} and
\ref{fig:top10country} display the top
ten ASes and countries from which we obtain \lhops;
for the hitlist-informed seed, 141 countries and 3,578 ASNs contribute
to the total.

\subsection{Edginess Metrics}
\label{sec:results:edginess}

To better understand the extent to which \edgy discovers 
\vsix \edge infrastructure, we introduce \replaced{three}{two} metrics of
``edginess.'' The first coarse metric is simply the
fraction of 
traces with a \lhop within the same \ac{AS} as the probe
destination.  Clearly, this condition does not imply that the \lhop
is truly an interface of the \edge router.  However, it provides a
rudimentary measure of whether traces are reaching the
target network's \ac{AS}.  In contrast, a trace to a non-existent
network will be dropped at an earlier hop in a default-free network.

We compare \edgy's results against
a day's worth of CAIDA's \vsix Ark traceroute results from
105 different vantage points on Oct 1, 2019~\cite{caida-topov6}.  Across nearly
17M traceroutes performed on that day, 1.7M (10\%) produced a
response from the target destination.  However, of those 1.7M traceroutes
that reached the destination, 86.2\% were from probing the 
\texttt{::1} address, while 13.3\% came from destinations known
to be aliased, \ie a fake reply.  Unsurprisingly, fewer than 0.5\%
of the probes to random targets reached the
\replaced{destination}{destinatino}.  

%

40.2\% of the CAIDA traces elicit a response from a \lhop address
that belongs to a BGP prefix originated by the same \ac{AS} as the
destination.  In contrast, 87.1\% of \edgy's traces reach the
target \ac{AS}.  While these results cannot be directly compared --
edgy performs two orders of magnitude more traces than CAIDA; see
\S\ref{results:caida} -- 
it does demonstrate that the probing performed by edgy is in fact
largely reaching the target network, if not the \edge. 


Our second edginess metric is a more granular measure of how deep into the
target network, and hence how close to the \edge, traces traverse.
For each trace, we find the number of most significant bits (MSBs)
that match between the target and the \lhop response, \ie 
the netmask of the most specific \vsix prefix that
encompasses the target and \lhop.  As before, this metric does not
provide a definitive measure of reaching the \edge.  Indeed, we
empirically observe many networks that use very different \vsix
prefixes for the \lhop point-to-point subnetwork as compared to the
customer's prefix.  However, the basis of this metric is that
hierarchical routing implies more matching
MSBs the closer the trace gets to the target.  

\begin{figure}[t]
 \centering
  \begin{minipage}{.5\textwidth}
    \centering
 \resizebox{1\columnwidth}{!}{\includegraphics[angle=-90]{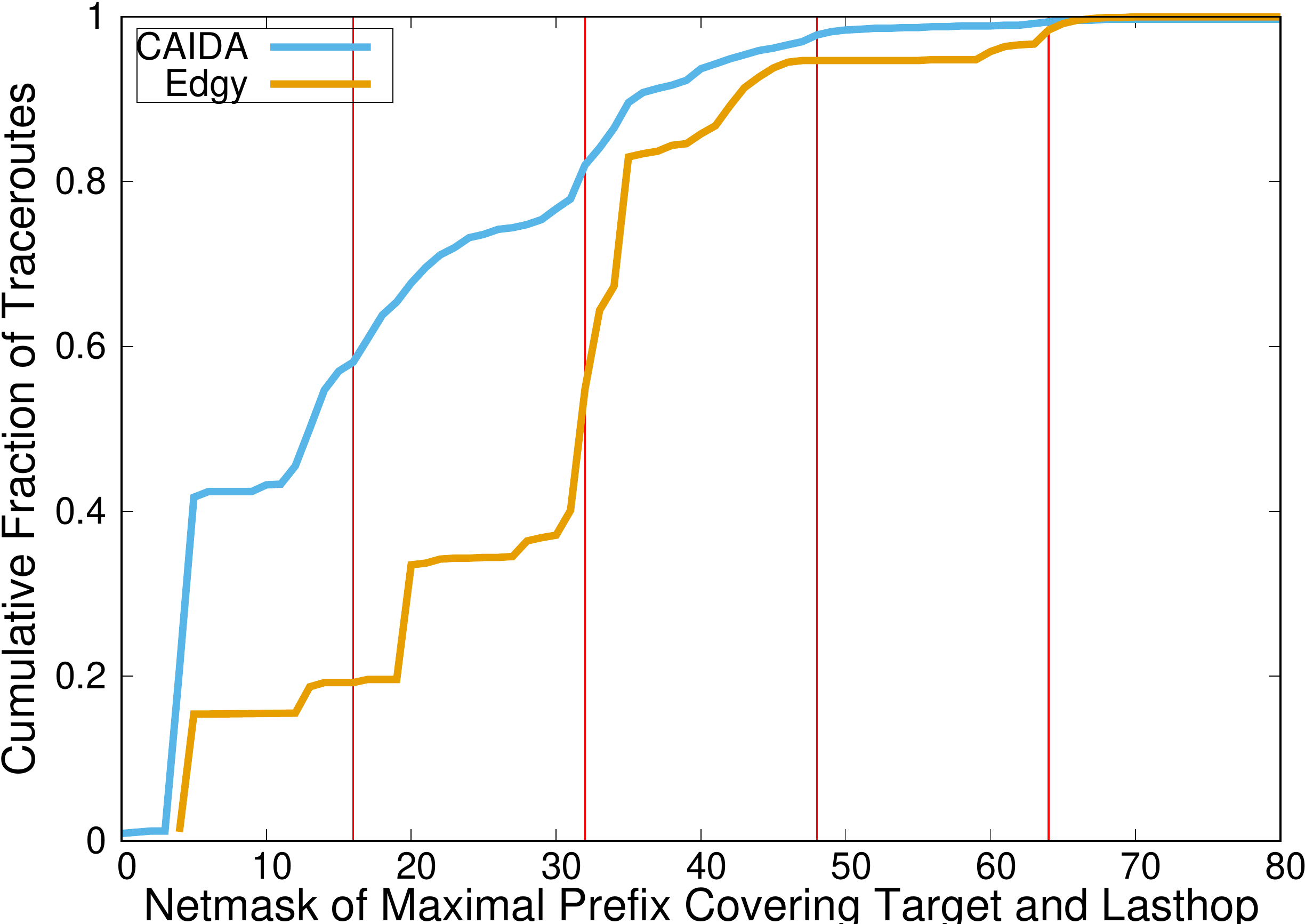}}
\vspace{-3mm}
  \caption{Size of prefix encompassing both target and \lhop \vsix addresses}
 \label{fig:dstbits}
  \end{minipage}%
  \begin{minipage}{.5\textwidth}
    \centering
\vspace{-8mm}
 \resizebox{1.1\columnwidth}{!}{\includegraphics{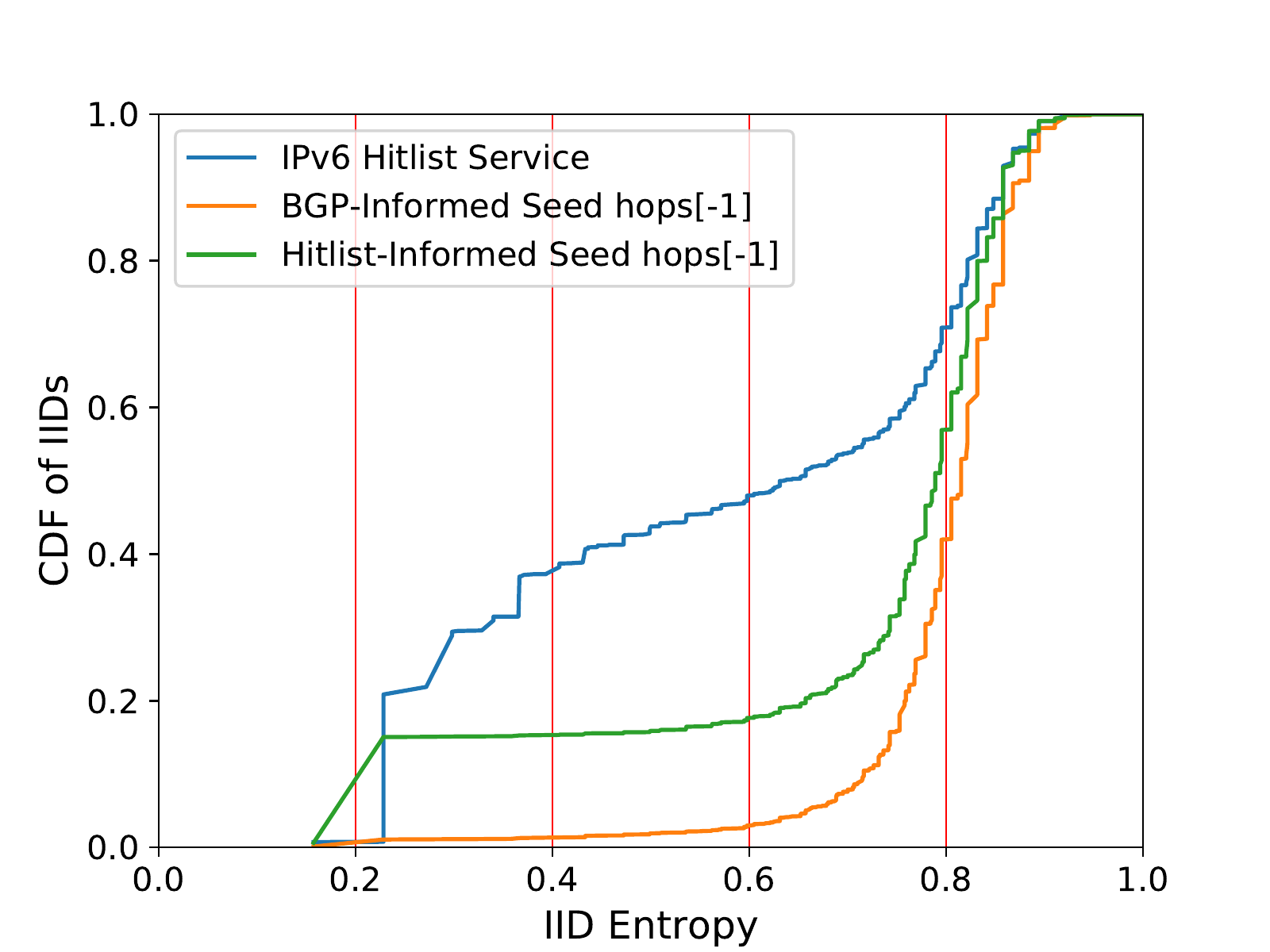}}
\vspace{-9mm}
  \caption{IID Entropies by Data Source}
 \label{fig:iid_entropy}
  \end{minipage}
\end{figure}

Figure~\ref{fig:dstbits} shows the distribution of matching bits
across the traceroutes from both CAIDA and \edgy.  Whereas the 
median size of the matching prefix is a /13 for CAIDA, it is
nearly a /32 for \edgy.  The target and \lhop share the same
/48 for more than 5\% of the \edgy traces, but just 2\% of the
CAIDA traces.  Thus, again, we see \edgy's probing reaching more
of the network \edge.

\added{Finally, we quantify how many of our \lhop addresses 
appear \emph{only} as periphery addresses in our traces, and therefore do not appear as an intermediate
hop in traceroutes to other target addresses. In the BGP-informed seed's first
round, 0.9\% of discovered last hop addresses to a target appear as an
intermediate hop to another target. In the second round, the same is true of
21\% of last hops, 23\% in the third round, and 4\% in the fourth probing round.
However, closer examination indicates that these numbers, particularly in the
second and third round, are skewed by providers
that frequently cycle periphery prefixes. For example, in the second round, 1.6M of the
2.5M addresses seen both as a last and an intermediate hop are located in
ASN8881, which we observe cycling customer prefixes on a daily
basis~\cite{rye2020discovering}. This often
causes traces to appear to ``bounce'' between two (or more) different addresses
toward the end of a trace. Sorting by the time the response was received
shows that a single \vsix address was responsible for high hop count responses
until after a distinct point at which a second address becomes responsive. This
erroneously causes the address that was not responsible for the highest hop
count response to appear as if it were an intermediate hop for the target.}

\added{We also observe a second class of \vsix address that appears both as a
\lhop and an intermediate hop to other targets.
These addresses appear as the last hop for a large number of target networks
that are most likely unallocated by the provider; these addresses typically have
low entropy \ac{IID} (\eg \ ::1 or ::2) and are likely provider
infrastructure.  These last hop addresses also appear on the path to addresses
that appear to be CPE, based on the high entropy or \eui last hop returned when
they are an intermediate hop.}

\subsection{Consolidated Results and Seed Data Comparison}
\label{sec:seed_comp}

Although both probing campaigns began with approximately
the same number of target /48 prefixes in the first probing round (130,447 and
111,670 in the BGP and hitlist-informed seeds, respectively), only 9,684 
/48s are common between the two data sets. The number of 
target prefixes in common decreases at each round, reaching 177 in 
/64 probing round.  Only $\sim$1.6M (2.5\%) \lhop \vsix addresses are
present in
both data sets. \added{These results demonstrate \edgy's sensitivity
to seed input,
and suggest that additional seed sources may aid discovery.}

Of the top ten ASNs, only four are common between
the two data sets -- ASNs 852, 8881, 45899, and 45609. Of the top ten countries, however,
six are common: Germany, Vietnam, Canada, Brazil,
India, and Japan, with Germany ranking first in both. While
the US is the second-leading producer of \lhop addresses in
the BGP-informed seed data with $\sim$6.9M unique \lhops, it is fourteenth in the
hitlist-informed data with only 357,877 addresses. Finally, we consider
the \lhop provider type 
using 
CAIDA's AS 
type classification~\cite{caida-classification}. 
By this classification, \edgy's results come
overwhelmingly from transit/access networks (99.9\%) rather than content or enterprise
ASes. 
This matches our
intent for \edgy to focus on \vsix \edge discovery.

\ifdefined\dox
\subsection{Vantage Point Sensitivity}

We conducted \edge discovery from a vantage point located in a California, USA
campus network in addition to our European vantage. Although both vantage points
were seeded with the same active discovery targets, the California campus
vantage point received significantly fewer \vsix addresses than the
European
vantage in the first probing round.  While the European vantage
discovered 4.6M
unique \lhops, the California monitor found only 3.4M. The 1.2M
address disparity between the two vantage points is due primarily to the
California vantage point obtaining fewer than 300 unique \lhops from Versatel
(ASN 8881), while our European vantage discovered more than 713,000.  Subsequent
experiments confirm our California vantage's inability to receive traceroute
replies from inside Versatel networks beyond its border, indicating filtering or
policy controls prohibiting \edge discovery in Versatel's network.  Excluding
heavy-hitter network anomalies, the California campus vantage point's first
round results largely follow those of the European vantage. However, because each
active network discovery round seeds the next, the effect of obtaining
significantly fewer \lhop addresses from the California vantage compounded at
each active discovery step.
\fi

\subsection{\eui Addresses}
\label{sec:results:eui64}

\begin{figure}[t]
 \centering
 \resizebox{.5\columnwidth}{!}{\includegraphics{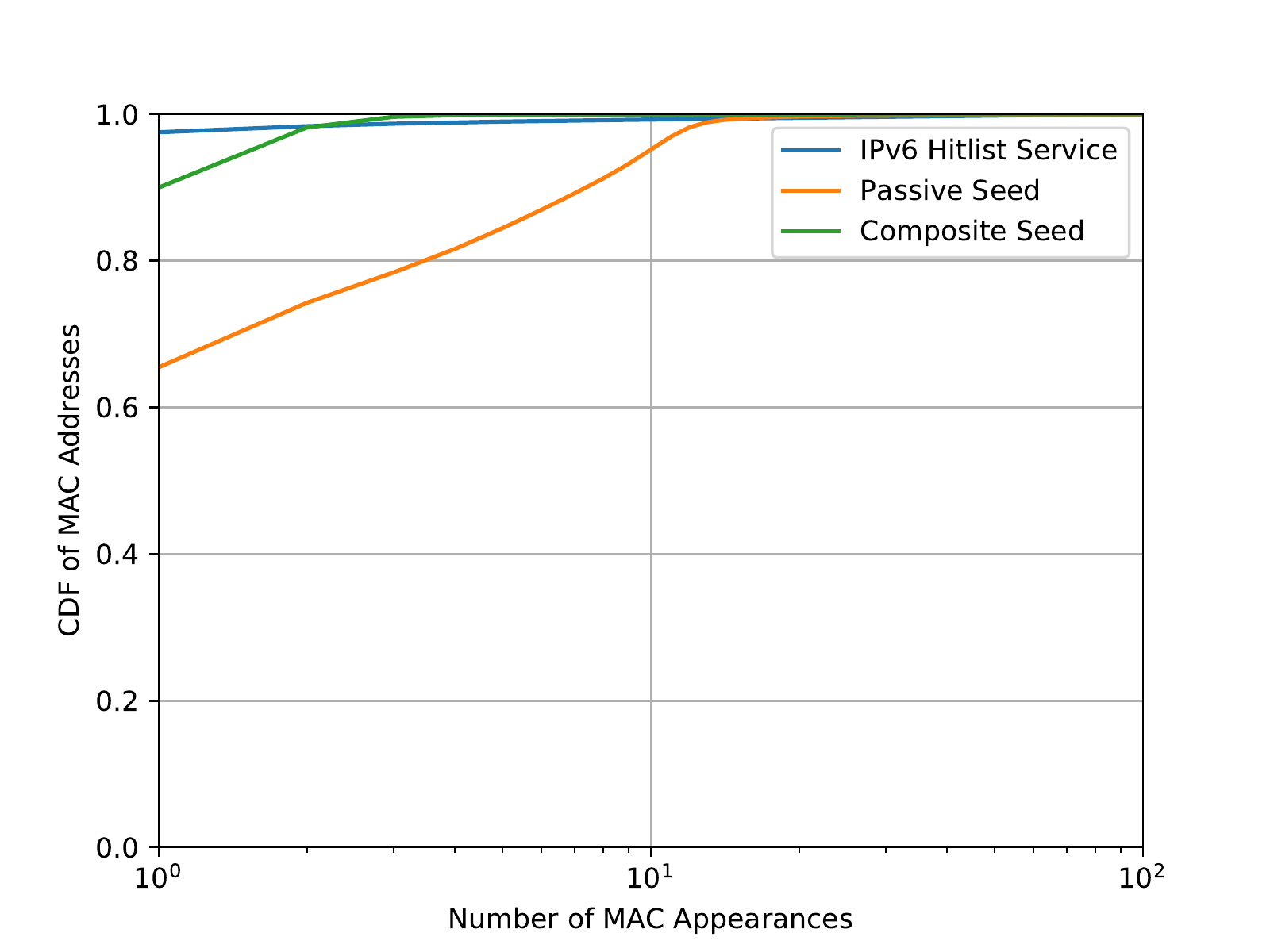}}
\vspace{-6mm}
  \caption{MAC Address Frequency by Source}
 \label{fig:mac_occur}
\end{figure}

Previous studies,
\eg~\cite{Gasser:2018:CEU:3278532.3278564,imc18beholder} identified
the presence of many \eui addresses in IPv6 traceroutes, where the
host identifier in the IPv6 address is a deterministic function of the
interface's \ac{MAC} address.  Our study similarly found a significant
fraction of \eui addresses, 
despite the introduction of privacy extensions for \ac{SLAAC} addresses in
2007~\cite{narten2007privacy}.
We discover slightly more than 16M \eui \lhop addresses, identifiable
from the \texttt{ff:fe} at byte positions 4 and 5 in an \ac{IID}, using the
BGP-informed 
seed data, or approximately 42\% of the total \lhops. 
However, only 5.4M (34\%) of the \ac{MAC} addresses in these
16M \lhops are unique.
Figure~\ref{fig:mac_occur} displays the CDF of
number of appearances MAC addresses make in unique \eui \lhop addresses.
65\% of the MAC addresses appear in only one \eui \vsix address;
30\% appear in two to ten \lhop addresses, and three MACs
appear in more than 10,000 different \lhops.

The discrepancy between unique \eui \lhop addresses and \ac{MAC} addresses appears to
have two root causes. The first is delegated prefix rotation.
Although 3.5M
of the 5.4M unique \ac{MAC} addresses observed appear
in only one \lhop address, 1.9M appear multiple times. Of these, the vast
majority appear in only several addresses in the same /48, suggesting that the
provider periodically rotates the remaining 16 bits of the network address
portion~\cite{zwangstrennung,rye2020discovering,rye2019eui64}. \added{We observe some providers rotating the
prefix delegated to their customers on a daily basis, and further examination of
forced prefix cycling is a topic of future work.} The second cause behind the disparity between number of \ac{MAC}
addresses and \eui \lhop addresses is due to what we believe is \ac{MAC} address
reuse.

For instance, the \ac{MAC} address \texttt{5\replaced{8}{A}:02:03:04:05:06}
occurs in more than 266k BGP-informed seed \lhop addresses in 76 /48s allocated to
providers throughout Asia and Africa. Because our probing took place over a
period of several weeks, we believe it is unlikely that a combination
of provider
prefix rotation and mobility substantially contributed to these;
its simple incremental pattern in bytes 2
through 6 further suggest it is likely a hard-coded \ac{MAC} address assigned to
every model of a certain device. \added{Support forums indicate that some models
of Huawei LTE router~\cite{dt-huawei,ru-huawei} use \texttt{58:02:03:04:05} as
an arbitrary \ac{MAC} address for their LTE WAN interface.}

\subsection{Comparison to the \vsix Hitlist Service}

We compare our results
to an open-source, frequently updated hitlist~\cite{Gasser:2018:CEU:3278532.3278564}.
In mid-October 2019, the
hitlist provides approximately 3.2M addresses responsive to \icmpsix, and
TCP and UDP probes on ports 80 and 443. 

Both the structure and magnitude of the addresses we discover
differentiate
our work
from~\cite{Gasser:2018:CEU:3278532.3278564}, which is unsurprising given our
focus on finding addresses at the network \edge. 
Unlike our results, the addresses in the hitlist are less likely to be \eui
addresses. Only $\sim$441,000 \eui addresses (with $\sim$338,000 unique MAC
addresses) appear in the hitlist, representing
approximately 14\% of the total responsive addresses. 
Figure~\ref{fig:iid_entropy} plots the normalized Shannon entropies of the \acp{IID} of
addresses in our datasets compared with addresses in the \vsix hitlist
service.
We see that the \vsix hitlist contains a far
greater proportion of low-entropy \acp{IID} addresses than the \lhop
addresses \edgy discovers. As \edge devices, particularly \ac{CPE} in residential
ISPs, are unlikely to be statically assigned a small constant \ac{IID} and instead
generate a high-entropy address via \ac{SLAAC}, this reinforces edgy's 
discovery of a different portion of the \vsix Internet
than prior work. Further emphasizing the complementary nature of
\edgy's probing, only 0.2\% of the
addresses we discover appear in this hitlist, indicating that \edgy discovers
different topology.  Finally, while the \lhops edgy
discovers overwhelmingly (99.9\%) reside in access networks (\S\ref{sec:seed_comp}),
CAIDA's AS-type classifier categorizes 1.8M of the hitlist's
\vsix addresses as residing in access/transit networks, 1.2M in content
networks, and 48k in enterprise networks.



\subsection{Comparison with CAIDA \vsix Topology Mapping}
\label{results:caida}

We again examine a day's worth
of CAIDA's \vsix Ark traceroute results from
105 different vantage points on Oct 1, 2019~\cite{caida-topov6}, 
to understand \edgy's 
complementary value.  
Because \edgy sends \replaced{nearly two orders of magnitude}{an
order of magnitude} more probes (544M vs 8.5M), these are not directly
comparable; however, we note that \edgy discovers 64.8M
non-aliased, routable \lhop addresses that CAIDA does not. 
CAIDA finds 163,952 unique, non-aliased, routable \lhop addresses. However,
despite focusing on only target networks that are dense in \lhops, \edgy still
discovers $\sim25\%$ of the \lhop addresses that CAIDA does. \Edgy similarly
finds 87.1M links to the \lhop address that CAIDA does not, but
discovers 54,024 of the 365,822 edges that contain only routable addresses from
CAIDA's probing. \added{Edgy's discovery of $\sim$37M unique \edge \lhops from
$\sim$544M targets probed in the BGP-informed seed yields 0.068 unique \lhops per
target, while the Ark traceroutes discover 0.019 unique \lhops
per target.}

\deleted{Of note, the disparity between the number of \lhop edges and
\lhops themselves is primarily due to addressing dynamics. While at first
glance the larger number of edges might suggest multihoming, because we believe
our discovered \lhops reside primarily on residential \ac{CPE}, we consider this
scenario unlikely. Considering all \lhop edges, the data reveals 17.1M
unique second-to-last hop \vsix addresses, with 59.3M \lhops. This
indicates that some second-to-last hops are responsible for a large number of
\lhops, and indeed, while 
99.9\% of the
unique last hops have ten
or fewer associated last hops, 14,593 second-to-last hops are part of the edge
to more than 100 \lhops. Thirty-four second-to-last hops appear in an edge to
over 100,000 \lhops, seventeen of which are AS8881, which we note
in~\cite{rye2020discovering} exhibits a high degree of prefix rotation among its
hosts.}

\subsection{Comparison with Seed Data Source}

\Edgy, by design, extends topology discovery
methodologies and is complementary to existing topology mapping campaigns.
However, because we believe \edgy provides increased
address discovery over existing mapping systems, we compare the results obtained
with \edgy to the trace seeds used as input to \edgy.

The BGP-informed seed source consists of traces conducted in August, 2018 to every
/48 in the routed \vsix Internet conducted from CAIDA's Archipelago~\cite{caida-routed48}. These
traces \added{to $\sim$711M unique targets} produce $\sim$5.8M 
unique last edges and 
$\sim$5.4M
unique \lhops after removing
non-routable addresses. By contrast, \edgy discovers
$\sim$59.5M 
unique final edges and 
$\sim$37.1M
unique \vsix \lhops \added{by probing to $\sim$545M targets when seeded with the
BGP-informed data}. Thus, \edgy significantly expands the discovered
topology of an input seed.

Likewise, \edgy discovers significantly more \lhop addresses and edges than the
hitlist-informed seed.
The hitlist-informed seed discovers
434,560 unique \lhops and 656,849 unique final edges, while \edgy, informed
by this data, discovers 
$\sim$29.2M
unique \lhops and 
$\sim$32.0M
final edges.

\ifdefined\dox
\subsection{Pathologies}
\label{sec:pathologies}

Versatel (ASN 8881), a German ISP, is the leading contributor of unique \vsix
\lhops in our BGP-informed seed data set, with 8,641,295 unique \lhop
addresses, 8,639,561 (99\%) of which are \eui. On closer examination, we find
that only 1,269,704 \ac{MAC} addresses occur in these \eui addresses\replaced{, implying
that on average each \ac{MAC} address appears in approximately seven \eui
addresses.}{; unique \ac{IID} occurrences range from 1 to 301 instances with a
median count of 7.}

This leads us to conclude that the unique \vsix address count for
AS8881 significantly over-counts \lhop equipment due to provider-enforced
prefix rotation, and suggests that addresses in AS8881 are particularly
ephemeral. 

Not all high address volume ASNs are caused by service providers re-delegating
prefixes at a higher-than-expected rate, however. For example, the
second-leading \vsix address contributor in our BGP-informed seed data set is
AS45899, VNPT Corp, a Vietnamese ISP. Of the 4,741,731 \vsix addresses in
AS45899, 925,257 of them are \eui and provide a stable \ac{IID} to track over
possible forced prefix rotations by the provider. In this case, 725,143 unique
\ac{MAC} addresses are present, implying that on average each \ac{MAC} address
appears in only 1.25 \vsix addresses. 
\fi

\section{Conclusions and Future Work}
We introduce \edgy, an algorithm to discover previously unknown portions of the
\vsix Internet, namely, the \vsix \edge. \Edgy extends
and augments existing \vsix discovery mapping systems, and the \lhop \edge
addresses that it discovers are nearly entirely disjoint from previous topology
mapping campaigns. \added{Because of privacy concerns involved with \eui
addresses and the ephemeral nature of many addresses, we are not 
releasing the periphery addresses \edgy discovers at this time; however, 
we expect our results to be reproducible.}

Several topics are planned for future work. First, we observe service
providers that cycle their customers' \edge prefix periodically.
This rotation leads to high levels of address discovery for these providers, but,
based on examining \ac{IID} reuse, over counts the number of actual device
interfaces present. We plan to: i) discover which networks implement
high-frequency prefix rotation; ii) quantify the rates at which new prefixes are issued;
and iii) determine whether the prefix issuing mechanism is
deterministic and predictable.
Second, we discover large numbers of \eui \vsix addresses more than a decade after
the introduction of \ac{SLAAC} privacy extensions~\cite{narten2007privacy}. Because \edgy discovers
\edge devices like \ac{CPE},
quantifying device types present in networks may be possible by
cross-referencing the models providers issue to customers, and through 
correlation with protocols that leak model information~\cite{acsac16furious}. 
Third, we wish to obtain more ground truth information on the \vsix
\edge as well as explicit validation of our results and algorithm.
\added{Fourth,} we plan to improve \edgy's efficiency by training it with
historical data \added{and leveraging multiple vantage points}. For
instance, \edge networks that exhibit frequent customer prefix cycling may need
to be probed on a regular basis, while those with stable \lhops
may be re-probed infrequently.
\added{Finally, because of the ephemeral nature of some of
the addresses we discover, we intend to couple other measurements
tightly with address discovery. For example, to further elucidate these
addresses' value, we will send \icmpsix Echo Requests and capture service
banners
immediately after receiving probe responses.}


\newpage
\section*{Acknowledgments}
\vspace{-1mm}
We thank Jeremy Martin, Thomas Krenc, and Ricky Mok for early feedback, John Heidemann
for shepherding, Mike Monahan and Will van Gulik for measurement
infrastructure, and the anonymous reviewers for insightful critique.
This work supported in part by NSF grant CNS-1855614.  Views and
conclusions are those of the authors and should not be interpreted as
representing the official policies or position of the U.S.\ government
or the NSF.

\small{
\bibliographystyle{splncs04}
\bibliography{edgy}

\begin{thebibliography}{10}
\providecommand{\url}[1]{\texttt{#1}}
\providecommand{\urlprefix}{URL }
\providecommand{\doi}[1]{https://doi.org/#1}

\bibitem{zwangstrennung}
Zwangstrennung (forced ip address change) (2018),
  \url{https://de.wikipedia.org/wiki/Zwangstrennung}

\bibitem{ru-huawei}
Huawei lte cpe b315 (mts 8212ft) - discussion (2019),
  \url{http://4pda.ru/forum/index.php?showtopic=700481&st=3580}

\bibitem{caida-classification}
{The CAIDA UCSD AS Classification Dataset} (2019),
  \url{http://www.caida.org/data/as-classification}

\bibitem{dt-huawei}
Speedport ii lte router status (2020),
  \url{https://telekomhilft.telekom.de/riokc95758/attachments/riokc95758/552/327892/1/routerstatus.pdf}

\bibitem{imc13dns}
Berger, A., Weaver, N., Beverly, R., Campbell, L.: {Internet Nameserver IPv4
  and IPv6 Address Relationships}. In: Proceedings of ACM Internet Measurement
  Conference (IMC) (2013)

\bibitem{imc16yarrp}
Beverly, R.: {Yarrp'ing the Internet: Randomized High-Speed Active Topology
  Discovery}. In: Proceedings of ACM Internet Measurement Conference (IMC) (Nov
  2016)

\bibitem{imc18beholder}
Beverly, R., Durairajan, R., Plonka, D., Rohrer, J.P.: {In the IP of the
  Beholder: Strategies for Active IPv6 Topology Discovery}. In: Proceedings of
  ACM Internet Measurement Conference (IMC) (Nov 2018)

\bibitem{caida-topov6}
{CAIDA}: {The CAIDA UCSD IPv6 Topology Dataset} (2018),
  \url{http://www.caida.org/data/active/ipv6_allpref_topology_dataset.xml}

\bibitem{caida-routed48}
{CAIDA}: {The CAIDA UCSD IPv6 Routed /48 Topology Dataset} (2019),
  \url{https://www.caida.org/data/active/ipv6_routed_48_topology_dataset.xml}

\bibitem{rfc7381}
Chittimaneni, K., Chown, T., Howard, L., Kuarsingh, V., Pouffary, Y., Vyncke,
  E.: {Enterprise IPv6 Deployment Guidelines}. RFC 7381 (Informational) (Oct
  2014), \url{https://www.rfc-editor.org/rfc/rfc7381.txt}

\bibitem{czyz2016}
Czyz, J., Luckie, M., Allman, M., Bailey, M.: {Don't Forget to Lock the Back
  Door! A Characterization of IPv6 Network Security Policy}. In: Network and
  Distributed Systems Security (NDSS) (2016)

\bibitem{Czyz:2014:MIA:2740070.2626295}
Czyz, J., Allman, M., Zhang, J., Iekel-Johnson, S., Osterweil, E., Bailey, M.:
  {Measuring IPv6 Adoption}. SIGCOMM Comput. Commun. Rev.  \textbf{44}(4) (Aug
  2014)

\bibitem{Dhamdhere:2012:MDI:2398776.2398832}
Dhamdhere, A., Luckie, M., Huffaker, B., claffy, k., Elmokashfi, A., Aben, E.:
  {Measuring the Deployment of IPv6: Topology, Routing and Performance}. In:
  Proceedings of ACM Internet Measurement Conference (IMC) (2012)

\bibitem{fan2010selecting}
Fan, X., Heidemann, J.: {Selecting Representative IP Addresses for Internet
  Topology Studies}. In: Proceedings of ACM Internet Measurement Conference
  (IMC) (2010)

\bibitem{Foremski:2016:EUS:2987443.2987445}
Foremski, P., Plonka, D., Berger, A.: {Entropy/IP: Uncovering Structure in IPv6
  Addresses}. In: Proceedings of ACM Internet Measurement Conference (IMC)
  (2016)

\bibitem{Gasser:2018:CEU:3278532.3278564}
Gasser, O., Scheitle, Q., Foremski, P., Lone, Q., Korczy\'{n}ski, M., Strowes,
  S.D., Hendriks, L., Carle, G.: {Clusters in the Expanse: Understanding and
  Unbiasing IPv6 Hitlists}. In: Proceedings of ACM Internet Measurement
  Conference (IMC) (2018)

\bibitem{rfc7707}
Gont, F., Chown, T.: {Network Reconnaissance in IPv6 Networks}. RFC 7707
  (Informational) (Mar 2016), \url{http://www.ietf.org/rfc/rfc7707.txt}

\bibitem{caida-ark}
Hyun, Y., k.~claffy: Archipelago measurement infrastructure (2018),
  \url{http://www.caida.org/projects/ark/}

\bibitem{rfc3177}
IAB, IESG: {Recommendations on IPv6 Address Allocations to Sites}. RFC 3177
  (Informational) (Sep 2001), \url{http://www.ietf.org/rfc/rfc3177.txt}

\bibitem{livadariu2015leveraging}
Livadariu, I., Ferlin, S., Alay, {\"O}., Dreibholz, T., Dhamdhere, A.,
  Elmokashfi, A.: {Leveraging the IPv4/IPv6 identity duality by using
  multi-path transport}. In: 2015 IEEE Conference on Computer Communications
  Workshops (2015)

\bibitem{Luckie:2017:IRO:3098822.3098858}
Luckie, M., Beverly, R.: {The Impact of Router Outages on the AS-level
  Internet}. In: Proceedings of ACM SIGCOMM (2017)

\bibitem{acsac16furious}
Martin, J., Rye, E.C., Beverly, R.: {Decomposition of MAC Address Structure for
  Granular Device Inference }. In: Proceedings of the Annual Computer Security
  Applications Conference (ACSAC) (Dec 2016)

\bibitem{Murdock:2017:TGI:3131365.3131405}
Murdock, A., Li, F., Bramsen, P., Durumeric, Z., Paxson, V.: {Target Generation
  for Internet-wide IPv6 Scanning}. In: Proceedings of ACM Internet Measurement
  Conference (IMC) (2017)

\bibitem{narten2007privacy}
Narten, T., Draves, R., Krishnan, S.: {Privacy Extensions for Stateless Address
  Autoconfiguration in IPv6}. RFC 4941 (Sep 2007),
  \url{http://www.ietf.org/rfc/rfc4941.txt}

\bibitem{rfc6177}
Narten, T., Huston, G., Roberts, L.: {IPv6 Address Assignment to End Sites}.
  RFC 6177 (Best Current Practice) (Mar 2011),
  \url{http://www.ietf.org/rfc/rfc6177.txt}

\bibitem{Plonka:2015:TSC:2815675.2815678}
Plonka, D., Berger, A.: {Temporal and Spatial Classification of Active IPv6
  Addresses}. In: Proceedings of ACM Internet Measurement Conference (IMC)
  (2015)

\bibitem{pujol2017understanding}
Pujol, E., Richter, P., Feldmann, A.: {Understanding the share of IPv6 traffic
  in a dual-stack ISP}. In: Passive and Active Measurement ({PAM}) (2017)

\bibitem{bcop-prefix}
RIPE: {Best Current Operational Practice for Operators: IPv6 Prefix Assignment
  for End-Users - Persistent vs Non-Persistent, and What Size to Choose}
  (2017), \url{https://www.ripe.net/publications/docs/ripe-690}

\bibitem{v6exhaust-ic16}
Rohrer, J.P., LaFever, B., Beverly, R.: {Empirical Study of Router IPv6
  Interface Address Distributions}. IEEE Internet Computing  (Aug 2016)

\bibitem{rye2020discovering}
Rye, E.C., Beverly, R.: {Discovering the IPv6 Network Periphery} (2020),
  \url{https://arxiv.org/abs/2001.08684}

\bibitem{rye2019eui64}
Rye, E.C., Martin, J., Beverly, R.: {EUI-64 Considered Harmful} (2019),
  \url{https://arxiv.org/pdf/1902.08968.pdf}

\bibitem{rfc2663}
Srisuresh, P., Holdrege, M.: {IP Network Address Translator (NAT) Terminology
  and Considerations}. RFC 2663 (Informational) (Aug 1999),
  \url{http://www.ietf.org/rfc/rfc2663.txt}

\bibitem{cymru2008ip}
{Team Cymru}: {IP to ASN mapping} (2019),
  \url{https://www.team-cymru.org/IP-ASN-mapping.html}

\bibitem{Zander:2018:WYI:3185332.3158374}
Zander, S., Wang, X.: {Are We There Yet? IPv6 in Australia and China}. ACM
  Trans. Internet Technol.  \textbf{18}(3) (Feb 2018)

\end{thebibliography}
}

\newpage
\begin{appendices}
\renewcommand{\thesection}{\appendixname~\Alph{section}}
\section{Algorithm Details}
\label{app:alg}
\vspace{-4mm}
\begin{algorithm}[]
\caption{Discover\_Init($seed\_traces$)}
\label{alg:discoverinit}
\begin{algorithmic}
 \State $density = []$
 \State $targets = []$
 \For {$(hops, dst) \in seed\_traces$}
   \State $dst48 \leftarrow dst$ \& $(2^{48}-1 \gg 80)$
   \State $LH \leftarrow hops[-1]$
   \State $density[LH] \leftarrow density[LH] \cup dst48$
 \EndFor
 \For {$LH \in density$}
   \If {$|LH| = 1$}
     \State $targets \leftarrow density[LH]$
   \EndIf
 \EndFor
  \For {$prefix \in targets$}
   \State Discover($prefix$)
   \EndFor
\end{algorithmic}
\end{algorithm}

\vspace{-5mm}
\begin{algorithm}[]
  \caption{Discover($prefix$)}
\label{alg:discover}
\begin{algorithmic}
  \State $masks = \{56, 60, 62, 64\}$
  \State $LH \leftarrow \{\}$
  \State $t \leftarrow rand(0,2^{64})$
  \For {$n \in masks$}
    \For {$i\leftarrow \{0\ldots 2^{n-48}-1\}$}
      \State $hops \leftarrow$ yarrp($prefix + (i \ll (128-n)) + t$)
      \State $LH \leftarrow hops[-1]$
    \EndFor
    \If {$|LH| \leq \eta$ or $n = 64$}
        \State $break$
    \EndIf
  \EndFor
\end{algorithmic}
\end{algorithm}

\ifdefined\dox
\vspace{-5mm}
\begin{algorithm}[]
\caption{Detection($dst$)}
\label{alg:detection}
\begin{algorithmic}
  \State $mask = 64$
  \Do 
    \State $LH \leftarrow \{\}$
    \State $t \leftarrow rand(0,2^{64})$
    \State $net \leftarrow dst$ \& $(2^{mask}-1 \gg 128-mask)$
    \State $hops \leftarrow$ yarrp($net - 2^{128-mask} + t$)
    \State $LH \leftarrow LH$ $\cup$ $hops[-1]$
    \State $hops \leftarrow$ yarrp($net + 2^{128-mask} + t$)
    \State $LH \leftarrow LH$ $\cup$ $hops[-1]$
    \State $mask \leftarrow mask + 1$
  \doWhile{$|LH| > 1$}
  \State return($mask$)
\end{algorithmic}
\end{algorithm}
\fi

\begin{figure}[t]
 \centering
  \begin{minipage}{.5\textwidth}
    \centering
 \resizebox{1\columnwidth}{!}{\includegraphics{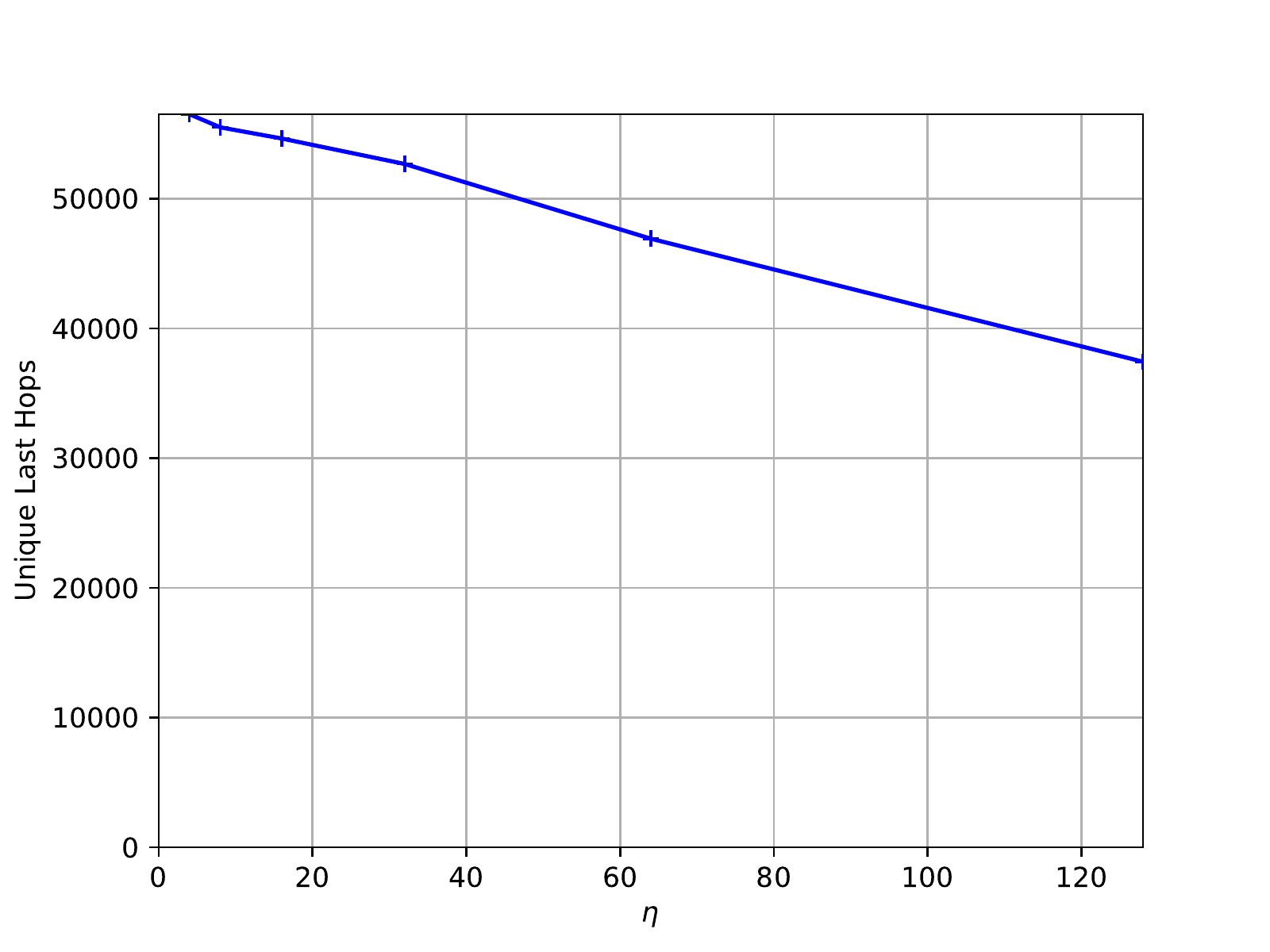}}
  \caption{Distinct \lhops for Selected $\eta$}
    \label{fig:eta_totalhops}
  \end{minipage}%
  \begin{minipage}{.5\textwidth}
    \centering
 \resizebox{1\columnwidth}{!}{\includegraphics{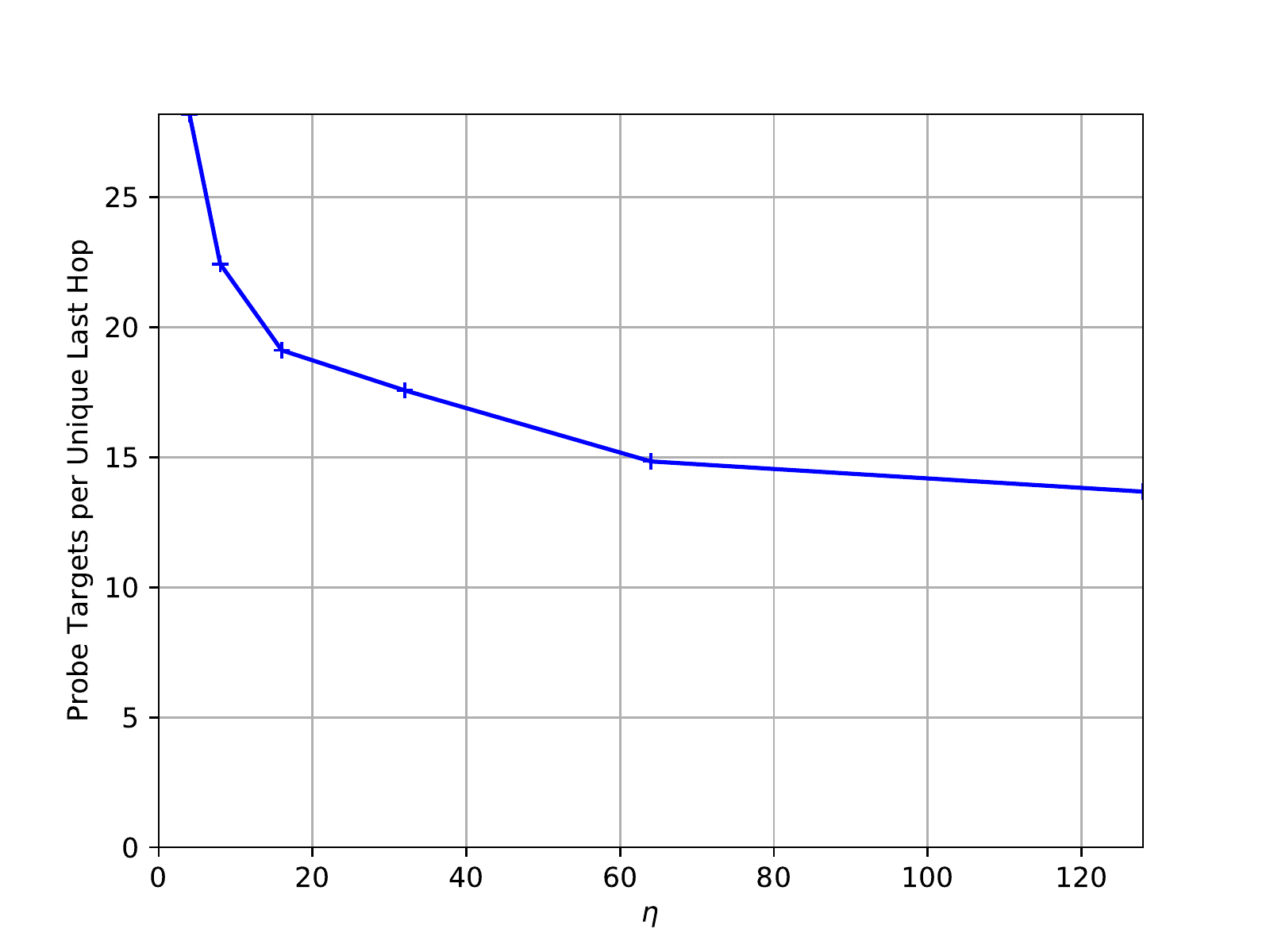}}
  \caption{Probe Efficiency for Selected $\eta$}
 \label{fig:eta_tgtsperhop}
  \end{minipage}
\end{figure}

\begin{table}[t]
  \centering
\begin{tabular}{|l|l|l|}
\hline
\textbf{$\eta$} & \textbf{Target Prefixes} & \textbf{Unique Last Hops} \\ \hline
4 & 389 & 56,507 \\ \hline
8 & 304 & 55,498 \\ \hline
16 & 255 & 54,636 \\ \hline
32 & 226 & 52,671 \\ \hline
64 & 170 & 46,913 \\ \hline
128 & 125 & 37,427 \\ \hline
\end{tabular}
  \caption{Sample Round Two Probing Results for Selected $\eta$}
  \label{table:sensitivity}
\end{table}

\section{Sensitivity Testing of $\eta$}
\label{app:eta}

In Algorithm~\ref{alg:discover}, we make use of the parameter $\eta$ in order to
determine whether or not to continue to probe a target prefix to attempt to
discover more \edge topology. Following the first probing round, we group unique
\lhops according to the target /48 prefix that produced them, and sort these 
target prefixes by number of unique \lhops. We wish to be judicious in the
prefixes selected to be probed in the second round of probing, selecting only
those prefixes that appear likely to return new \edge addresses, balancing the
amount of time incurred by each additional prefix selected, and leveraging the
best current practice to subnet on nybble boundaries~\cite{bcop-prefix}. For
the last reason, we initially choose $\eta=16$; the first round exhaustively
probes all /56s inside of a /48, and obtaining 17 or more unique last hop
addresses is indicative of subnetting below the /52 level. We seek to validate
this choice empirically, however, and as such conduct sensitivity analysis on a
sample of target prefixes from the BGP-informed seed data.

We randomly sample 1,000 prefixes from the BGP-informed seed data and use \yarrp to
trace to a random \ac{IID} in each /56 of every target /48 prefix. All but one
of the target prefixes obtains at least one unique \lhop. From here, we
choose $\eta \in \{4,8,16,32,64,128\}$, select the target prefixes with $i >
\eta$ distinct \lhops, and conduct the second round of probing using these
selected prefixes. Table~\ref{table:sensitivity} summarizes the number of target
prefixes selected for each choice of $\eta$, and as well as the count of unique
\lhops obtained using each $\eta$. Figure~\ref{fig:eta_totalhops} displays the
number of unique \lhops for each choice of $\eta$ graphically. In order to
determine which $\eta$ choice is most efficient in unique \lhop discovery, we
plot the number of probe targets divided by the number of distinct \lhops
discovered at each $\eta$. Of
the values chosen, $\eta=16$ clearly lies at an inflection point in
Figure~\ref{fig:eta_tgtsperhop}, indicating that while each successive $\eta$
value is more efficient, efficiency increases at a decreasing rate. Because of
the efficiency of $\eta=16$, combined with its high absolute address discovery,
we choose $\eta=16$ as the threshold for passage to round 2 of the address
discovery stage.
\end{appendices}

\begin{acronym}
  \acro{AS}{Autonomous System}
  \acro{CAIDA}{Center for Analysis of Internet Data}
  \acro{CPE}{Customer Premises Equipment}
  \acro{EUI}{Extended Unique Identifier}
  \acro{FPGA}{Field-Programmable Gate Array}
  \acro{IID}{Interface IDentifier}
  \acro{MAC}{Media Access Control}
  \acro{SOHO}{Small Office-Home Office}
  \acro{SLAAC}{Stateless Address Autoconfiguration}
\end{acronym}

\end{document}